\documentclass[conference]{IEEEtran}
\IEEEoverridecommandlockouts

\usepackage{cite}
\usepackage{amsmath,amssymb,amsfonts}
\usepackage{algorithmic}
\usepackage{graphicx}
\usepackage{textcomp}
\usepackage{xcolor}
\usepackage{hyperref}
\usepackage{siunitx}
\usepackage{fancyvrb}
\usepackage{siunitx}
\usepackage{hyperref}
\usepackage{tikz}

\usepackage{xcolor}

\usepackage{lipsum}
\usepackage[pscoord]{eso-pic}
\newcommand{\placetextbox}[3]{
  \setbox0=\hbox{#3}
  \AddToShipoutPictureFG*{
    \put(\LenToUnit{#1\paperwidth},\LenToUnit{#2\paperheight}){\vtop{{\null}\makebox[0pt][c]{#3}}}%
  }%
}%

\def\BibTeX{{\rm B\kern-.05em{\sc i\kern-.025em b}\kern-.08em
    T\kern-.1667em\lower.7ex\hbox{E}\kern-.125emX}}
\begin{document}
\placetextbox{0.5}{1}{This is the author's version of an article that has been published.}
\placetextbox{0.5}{0.985}{Changes were made to this version by the publisher prior to publication.}
\placetextbox{0.5}{0.97}{The final version of record is available at \href{https://doi.org/10.1109/ETFA65518.2025.11205552}{https://doi.org/10.1109/ETFA65518.2025.11205552}}%
\placetextbox{0.5}{0.05}{Copyright (c) 2025 IEEE. Personal use is permitted.}
\placetextbox{0.5}{0.035}{For any other purposes, permission must be obtained from the IEEE by emailing pubs-permissions@ieee.org.}%

\title{QRmap: executable QR codes for Navigation in Industrial Environments and Beyond\vspace{-0.3cm}\\
\thanks{This work has been partially funded by CNR under the project ``Executable QR codes (EQR) - GORU IEIIT'' (DIT.AD001.212).}
}

\author{
    \IEEEauthorblockN{
    Stefano Scanzio\IEEEauthorrefmark{1},
    Paolo Campagnale\IEEEauthorrefmark{2},
    Pietro Chiavassa\IEEEauthorrefmark{1},
    Gianluca Cena\IEEEauthorrefmark{1}
    }
    
    \IEEEauthorblockA{\IEEEauthorrefmark{1}National Research Council of Italy (CNR--IEIIT), Italy. \IEEEauthorrefmark{2}Politecnico di Torino, Italy.}
    Emails: \{stefano.scanzio, pietrochiavassa, gianluca.cena\}@cnr.it, paolocampagnale@gmail.com\vspace{-0.3cm}
    }

\maketitle

\begin{abstract}
QR codes are nowadays customarily used for embedding static data such as web hyperlinks or plain text. 
The sQRy technology (executable QR codes) permits to embed executable programs in QR codes, enabling people to interact with them even without an 
internet connection.

In this work we present QRmap, a specific dialect that permits the inclusion of geographic maps in sQRy and supports interaction with the user to provide indications to reach the destination of interest. 
The QRmap technology facilitates navigation in large industrial plants where internet connectivity is absent, due to either environmental limitations or company policies. 
The proposed technology can have interesting applications in non-industrial contexts as well. 
\end{abstract}

\section{Introduction}
Since their introduction in 1994, QR codes have been used in an increasing number of application contexts, including education~\cite{10619878}, smart documents~\cite{10594158}, digital payments~\cite{10698108}, security~\cite{IoT1}, quality insurance~\cite{15}, and other~\cite{10492995}. 
In all these cases, QR codes contain a static identifier, which is typically a uniform resource identifier (URI) that redirects to a server-side application. 
The possibility to embed executable programs in QR codes came later, in 2022~\cite{9921530}, with the invention of sQRy (\url{https://www.sqry.org}), which are also known as executable QR (eQR) codes. 
sQRy supports user interaction in the absence of a reliable internet connection
because the program (i.e., the application) is completely stored in the QR code itself.

One of the most severe limitations of the QR code technology is storage capacity~\cite{scanzio2025compressionexecutableqrcodes}, which reaches a maximum of $\SI{2953}{bytes}$ in the most capacious version 40 with a low error correction level (that is coded as a $177\times 177$ matrix). 
Choosing a low error correction decreases robustness in worn QR codes.
Depending on the intended application, different languages/dialects have been defined, each one with specific trade-offs between expressiveness and space occupation. 
Possible application contexts of sQRy include industrial environments characterized by poor connectivity, such as those located in desert areas, mountains, and petrochemical plants, as well as situations where company policies do not allow users to set up network connections, e.g., for security guidelines.

In the context of this work, we propose a new dialect for navigation called QRmap. The geographical map or floor plan of a factory is coded in the sQRy. 
In addition, QRmap defines a programming language that interacts with the user, answering questions regarding the reachable destinations and guiding the user to the intended location by recommending the path with the shortest travel time or distance. The dialect was designed to represent any kind of graphs and their related environments.
QRmap could have a significant impact in many industrial application contexts, where different operators (e.g., employees, maintenance technicians, visitors, healthcare personnel, installation specialists, etc.) 
can be guided to various destinations throughout the industrial plant, including muster points, emergency exits, and infirmaries. 
For instance, a technician who needs to locate a specific machine or component within a production cell can obtain instructions on how to reach the correct destination by means of a sQRy containing a QRmap. 
Similarly, people in the factory can be conducted to the nearest infirmary or emergency exit, even in the absence of a network connection or during a black-out. 
Navigation can be performed in multiple steps: 
a first sQRy guides the selection of a specific building within the whole industrial plant, while a second sQRy guides the user to a destination inside the building. 
Of course, QRmap usage is not limited to industrial settings.

QR codes or other tags such as RFID have already been used for navigation, but in these contexts they only store an identifier to communicate to a robot or a machinery the position. On the contrary, sQRy based on QRmap are mainly meant for, but not limited to, human operators, and they permit an interaction to detect the intended destination and the best path given the user's requirements.

The next Section~\ref{sec:sQRy} of this work introduces the sQRy technology, which is needed for the definition of the QRmap dialect for navigation in Section~\ref{sec:QRmap} and for the commented application example of Section~\ref{sec:example}. Finally, conclusive remarks are reported in Section~\ref{sec:conclusions}.

\section{sQRy / eQR codes}
\label{sec:sQRy}
\begin{figure}[b]
\begin{center}
\vspace{-0.5cm}
\includegraphics[width=1.0\linewidth]{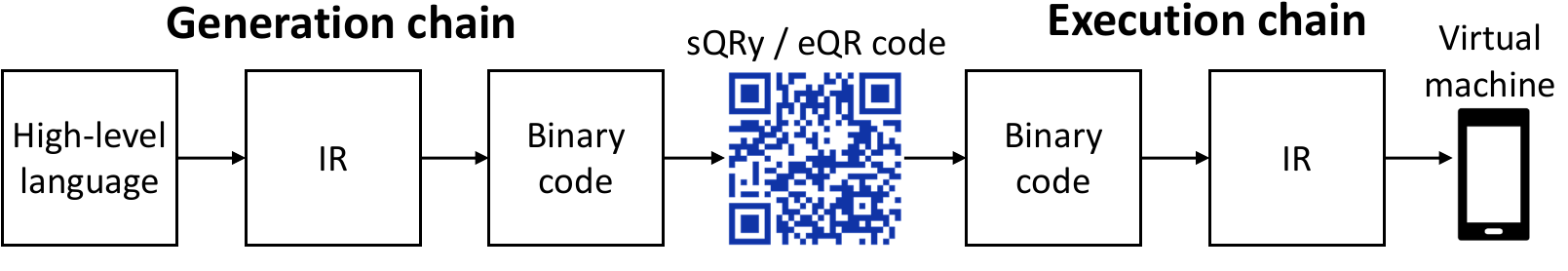}
\end{center}
\vspace{-0.3cm}
\caption{sQRy generation and execution chains.}
\label{fig:chain}
\end{figure}
A specific \textit{generation} chain that produces the sQRy (i.e., a QR code embedding an executable program) and an \textit{execution} chain that executes the sQRy on a device (e.g., smartphone, browser, QR code reader) are required and sketched in Fig.~\ref{fig:chain}.

Concerning the \textit{generation} chain, a high-level programming language, either textual or graphical, is translated into an intermediate representation (IR) composed of simple instructions. 
Three-address code is a common example of IR derived from compilers' theory.
In the case of sQRy, IR instructions are defined at a higher level because they are not intended to be directly executed by a CPU, but by a virtual machine running on relatively fast devices. 
The IR is then translated into a binary format that can be embedded in standard QR codes using open-source libraries. 
IR relies on a standardized set of instructions, which can be used by a multitude of high-level languages and provide the ability to efficiently code a specific dialect in a binary format. 
The most important (and complex) part of the definition of any new dialect, including QRmap, is the definition of the instruction set for IR and the way they are converted into a binary representation and vice versa, with particular emphasis on the compactness of the generated binary code.

On the other hand, the \textit{execution} chain is used for executing the embedded program. 
Specific and widely available libraries can be exploited to acquire and decode the QR code image, which contains the binary representation. 
This binary data is then translated back into an IR, which is executed by a specific virtual machine on the end device (typically a smartphone).

Two sQRy dialects are currently defined: 
QRtree for the execution of decision tree~\cite{10492995,QRscript-spec,QRtree-spec}, which is characterized by very good compactness (i.e., small binary code), 
and QRind~\cite{10710739} for industrial applications, which also permits to embed simple machine learning algorithms. 
An open-source implementation of QRtree is available~\cite{QRtree-soft}.

\section{The QRmap dialect}
\label{sec:QRmap}
The QRmap dialect permits to encode geographic maps into a sQRy, enabling users to determine the best path to the destination they wish to reach.
QRmap instructions, which are embedded in the sQRy header, are composed of five sections in the following order: header, label, graph, quick choice, and code.

The \textit{graph section} is used to map physical spaces into a graph representation $G=(V, E)$, which is composed of a set of vertex $V=(v_1,v_2,...,v_{|V|})$ and a set of arcs or edges $E=(e_1,e_2,...,e_{|E|})$, where $|V|$ and $|E|$ are the number of vertices and edges, respectively. 
Each edge connects two vertices. Both directed and undirected arcs can be represented in QRmap. 
With the notation $e_{v_i \rightarrow v_j}$, we represent an arc between vertices $v_i$ and $v_j$; 
instead, a possible notation for an undirected arc is $e_{v_i \leftrightarrow v_j}$. 
Vertices can be used to model parts of a plant, machinery, or rooms. 
Instead, arcs are used to model distances (or any other weighting function, e.g., energy consumption) by means of labels that serve to describe a segment of a path.

A \textit{label} can be associated with arcs and vertices. 
Let $e_{i}^{l_j}$ denote arc $i$, and $l_j \in L$ represents the label associated with this arc.
$L=\{ l_1,l_2,...,l_{|L|} \}$ is the set of all the possible labels that can be assigned to arcs and vertices. 
These labels, whose number is $|L|$, are defined in the label section. 
A similar notation is defined for vertices. 
A set of attributes can be associated with every label, so we can write
$l_j = {\{\mathrm{.a}_1,\mathrm{.a}_2,...,\mathrm{.a}_{|l_j|}\}}$ 
to specify the $|l_j|$ attributes associated to label $l_j$. 
For example, 
$l_1={\{\mathrm{.name=``hop"}, \mathrm{.n=3} \}}$ means that label $l_1$ has two attributes: 
\texttt{.name}, whose value is set to the string \texttt{"hops"}, and  
\texttt{.n}, whose value is set equal to the integer number \texttt{3}. 
Notation $e_1^{\{\mathrm{.name=``hops"}, \mathrm{.n=3} \}}$ expresses how label attributes are valorized for arc $e_1$.

Labels permit to separate the values assumed by attributes from their description. 
In order to save space, the description is made only once, and the association of labels with arcs and vertices is made in a positional way. 
There is the possibility of single inheritance between labels. 
A label can extend the attributes inherited from another label with its own attributes.
The \textit{label section}, which is defined before the graph section, is used to define labels.

All the other sections are optional. 
The \textit{header section}, which is the first, starts with the command \texttt{QRMAP\_HEADER}, which is followed by a list of $13$ commands used to define and calibrate some features for the following sections. 
Possible settings regard the coding of integer and floating-point numbers, or the measurement units used, for instance, to define distances or the time to travel along an arc.

The \textit{quick choices section} is located after the graph section. 
A quick choice is a complex search criterion that defines the search parameters for a route to a specific destination. 
Search criteria may contain a set of requirements on the nodes and arcs traversed, which can be connected using Boolean expressions. 
To avoid the use of parentheses and save memory, expressions are stored in reverse polish notation. 
These conditions will be rendered in the mobile application on the user device with buttons that, when pressed, apply the defined search criteria and start the navigation to the intended location. 
Among other things, quick choices can include, e.g., selecting the nearest emergency exit, bathroom, or infirmary.

\begin{figure}[t]
\begin{center}
\includegraphics[width=0.90\linewidth]{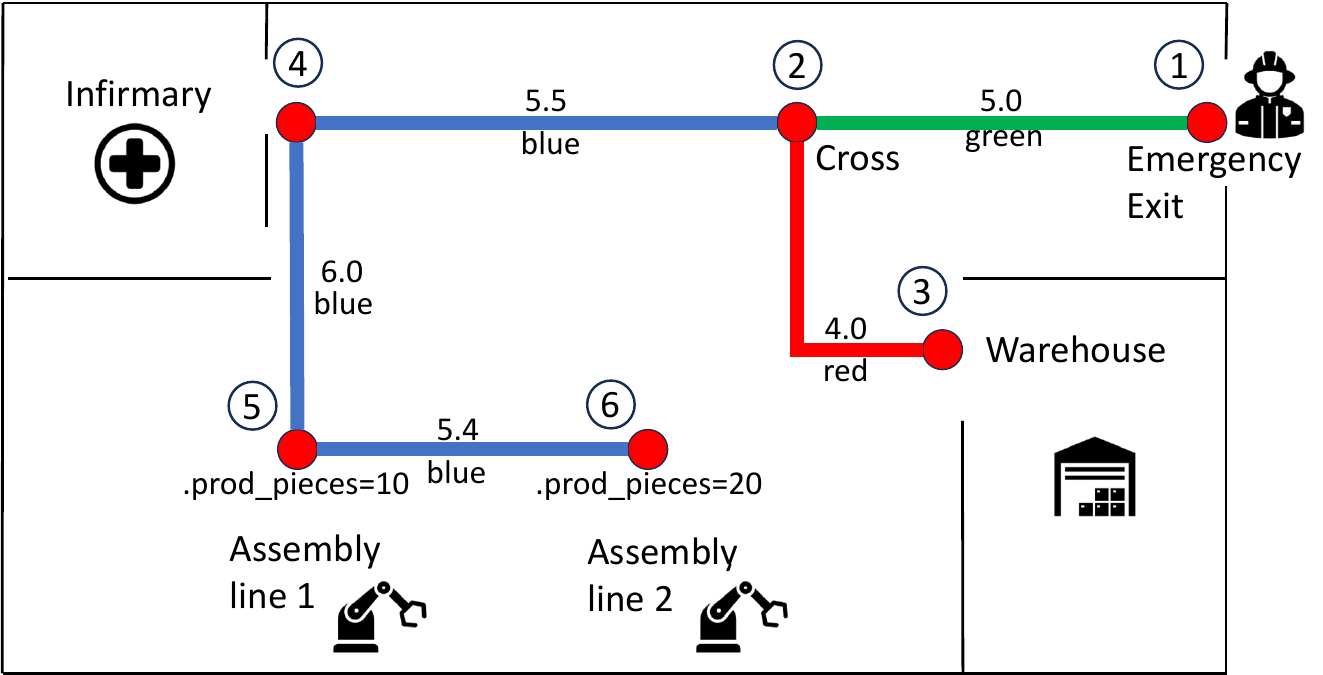}
\end{center}
\vspace{-0.3cm}
\caption{Example of a floor plan in which to use QRmap dialect.}
\vspace{-0.3cm}
\label{fig:layout}
\end{figure}

The last \textit{code section} contains the instructions for the user organized as a decision tree. 
After asking the user a number of questions, the application feeds her/his answers to a path-finding algorithm, which determines the optimal path to reach the destination. 
Algorithms for finding optimal paths on graphs are well-documented in scientific literature and were therefore not detailed in this paper.
Security aspects were not discussed in this work, but they can be addressed with traditional methods based on public/private keys for authentication.

\section{Example}
\label{sec:example}
A simple, concrete example is provided to illustrate a practical application scenario where QRmap can be profitably exploited, along with the related implementation details.

In Fig~\ref{fig:layout}, the floor plan of a simplified (large) industrial plant is depicted. There are six identified points of interest, reachable through the sQRy based on QRmap: 
an emergency exit, an intersection that permits reaching the following two points of interest (a warehouse and an infirmary), and two assembly lines. 
Assembly line 1 can produce $10$ pieces of a given good every hour, while assembly line 2 is faster and can produce $20$ pieces per hour. 
In the sample industrial plant, the paths connecting the points of interest are painted on the floor with specific colors, which are used to differentiate the different routes. 
Another possibility is to number them. 
In addition, a distance (in meters) is associated with every path.

The decision to analyze a simplified plant with a reduced number of vertices (i.e., points of interest) is aimed at making the QRmap code (reported in the following) compact and easy to analyze by the reader.

\begin{figure}[t]
\begin{center}
\includegraphics[width=1.00\linewidth]{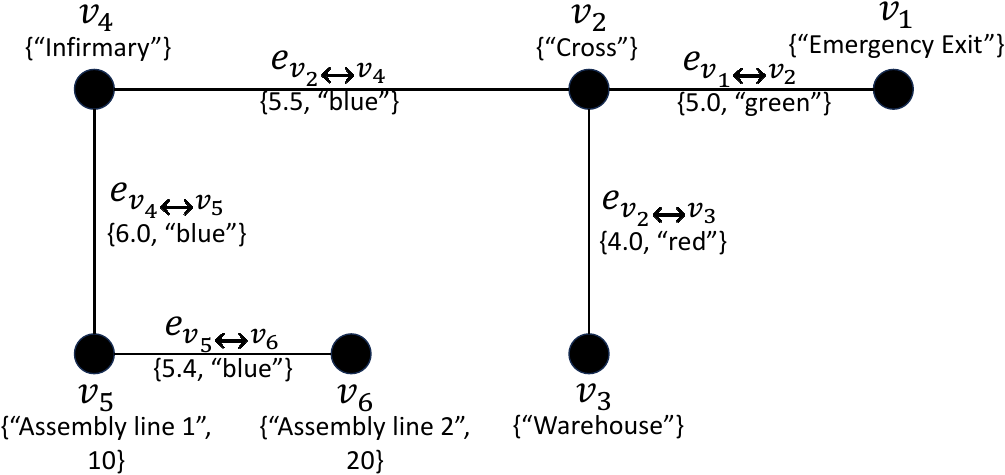}
\end{center}
\vspace{-0.3cm}
\caption{Graph representation of the floor plan reported in Fig.~\ref{fig:layout}, which can be coded through QRmap.}
\vspace{-0.4cm}
\label{fig:graph}
\end{figure}

The floor plan can be translated into the graph representation of Fig.~\ref{fig:graph}.
In this example, three labels are defined: 
$l_1^{\{\mathrm{.name},\mathrm{.desc}\}}$ with attributes \textit{name} and \textit{desc} (i.e., name and description of a vertex), 
$l_2^{\{\mathrm{.length},\mathrm{.color}\}}$ with attributes \textit{length} and \textit{color} (i.e., the length and the color of an arc), and
$l_3^{\{l_1, \mathrm{.prod\_pieces}\}}$ with the attributes inherited by $l_1$ plus the attribute \textit{prod\_pieces} (i.e., name, description, and produced pieces of a vertex).

Valorization of the attributes associated with vertices and arcs 
is also reported in the figure.
In the example, $v_3^{l_1}=v_3^{\{\mathrm{.name=``Warehouse"}\}}$ represents a warehouse vertex, where the \textit{desc} attribute was left empty because, as will be illustrated in the QRmap intermediate language example of Fig.~\ref{fig:intermediate}, it is optional. 
Instead, for assembly line 2, $v_6^{l_3}=v_6^{\{\mathrm{.name=``Assembly\ line\ 2"},\mathrm{.desc=``Fast"},\mathrm{.prod\_pieces=20}\}}$. 
Arc $e_{v_1 \leftrightarrow v_2}^{l_2}=e_{v_1 \leftrightarrow v_2}^{\{\mathrm{.length=5.0},\mathrm{.color=``green"},\mathrm{.prod\_pieces=20}\}}$ models instead a segment of path colored in green whose length is $5.0$ meters.

\begin{figure*}[t]
\fontsize{7}{7}\selectfont
\linespread{0.85}
\begin{Verbatim}[commandchars=\\\{\}]
/* a)\textbf{ QRmap header section} */
\textbf{QRMAP_HEADER}; ...; \textbf{DEFINE_UNIT} VALUE ASCII7 \textbf{"m"}; \textbf{ENCODE_FLOAT} 3 4; ...

/* \textbf{b) Label section} */
\textbf{DEFINE_LABEL} NAME ASCII7 \textbf{"common"} STRING ASCII7 \textbf{"name"} NO_OPTIONAL STRING ASCII7 \textbf{"desc"} OPTIONAL NO_DEFAULT
\textbf{DEFINE_LABEL} NAME ASCII7 \textbf{"arc"} FLOAT ASCII7 \textbf{"length"} MEASUREMENT_UNIT 0 NO_OPTIONAL STRING ASCII7 \textbf{"color"} NO_OPTIONAL
\textbf{DEFINE_LABEL} NAME ASCII7 \textbf{"ass_line"} EXTENSION 1 INT ASCII7 \textbf{"prod_pieces"} MEASUREMENT_UNIT 0 NO_OPTIONAL

/* \textbf{c) Graph section} */
\textbf{LABEL 1} ATTRIBUTES_FLAGS TRUE ATTRIBUTES_VALUES STRING ASCII7 \textbf{"Emegency Exit"} STRING ASCII7 \textbf{"Use in case of emergency"}
    \textbf{DESTINATION_NODE 2} NO_ORIENTED \textbf{LABEL 2} ATTRIBUTES_FLAGS FALSE ATTRIBUTES_VALUES FLOAT \textbf{5.0} STRING ASCII7 \textbf{"green"}
\textbf{LABEL 1} ATTRIBUTES_FLAGS FALSE ATTRIBUTES_VALUES STRING ASCII7 \textbf{"Cross"} 
    \textbf{DESTINATION_NODE 3} NO_ORIENTED \textbf{LABEL 2} ATTRIBUTES_FLAGS FALSE ATTRIBUTES_VALUES FLOAT \textbf{4.0} STRING ASCII7 \textbf{"red"}
    \textbf{DESTINATION_NODE 4} NO_ORIENTED \textbf{LABEL 2} ATTRIBUTES_FLAGS FALSE ATTRIBUTES_VALUES FLOAT \textbf{5.5} STRING ASCII7 \textbf{"blue"}
\textbf{LABEL 1} ATTRIBUTES_FLAGS FALSE ATTRIBUTES_VALUES STRING ASCII7 \textbf{"Warehouse"}
\textbf{LABEL 1} ATTRIBUTES_FLAGS FALSE ATTRIBUTES_VALUES STRING ASCII7 \textbf{"Infirmary"}
    \textbf{DESTINATION_NODE 5} NO_ORIENTED \textbf{LABEL 2} ATTRIBUTES_FLAGS FALSE ATTRIBUTES_VALUES FLOAT \textbf{6.0} STRING ASCII7 \textbf{"blue"}
\textbf{LABEL 3} ATTRIBUTES_FLAGS TRUE ATTRIBUTES_VALUES STRING ASCII7 \textbf{"Assembly line 1"} STRING ASCII7 \textbf{"Slow"} INT \textbf{10}
    \textbf{DESTINATION_NODE 6} NO_ORIENTED \textbf{LABEL 2} ATTRIBUTES_FLAGS FALSE ATTRIBUTES_VALUES FLOAT \textbf{5.4} STRING ASCII7 \textbf{"blue"}
\textbf{LABEL 3} ATTRIBUTES_FLAGS TRUE ATTRIBUTES_VALUES STRING ASCII7 \textbf{"Assembly line 2"} STRING ASCII7 \textbf{"Fast"} INT \textbf{20}


/* \textbf{d) Quick choices section} */
\textbf{DEFINE_QUICK_CHOICES} QUICK_CHOICE SHORTEST_PATH ASCII7 \textbf{"length"} ID_LABEL_NE NODE \textbf{DESTINATION 1} 

/* \textbf{e) Code section} */
(0) \textbf{INPUT} ASCII7 "What is your destination?"
(1) \textbf{IF} ASCII7 "Emergency" (4)
(2) \textbf{IF} ASCII7 "Assembly line" (6)
(3) \textbf{GOTO} (15)
(4) \textbf{SEARCH} QUICK_CHOICE 0
(5) \textbf{GOTO} (15)
(6) \textbf{INPUTS} ASCII7 "How many products do you want to produce in an hour?"
(7) \textbf{IFC} <= INT 0 3 10 (12)
(8) \textbf{PUSH} ID_NE NODE DESTINATION 6
(9) \textbf{PRINT} "You are directed to Assembly line 2"
(10) \textbf{SEARCH} GUIDED_SEARCH SHORTEST_PATH ASCII7 "length"
(11) \textbf{GOTO} (15)
(12) \textbf{PUSH} ID_NE NODE DESTINATION 5
(13) \textbf{PRINT} "You are directed to Assembly line 1"
(14) \textbf{SEARCH} GUIDED_SEARCH SHORTEST_PATH ASCII7 "length"
\end{Verbatim}
\vspace{-0.1cm}
\caption{Intermediate representation of the QRmap code included in the sQRy.}
\vspace{-0.4cm}
\label{fig:intermediate}
\end{figure*}

A program based on the QRmap IR language, which assists users when they move in the floor plan of Fig.~\ref{fig:layout}, is reported in Fig.~\ref{fig:intermediate}.
We remark again that the IR language is typically obtained automatically from a high-level programming language, which is why it is often more difficult to read. 
However, the definition of the IR language is very important, as it is the part that needs to be standardized and has to be considered, therefore, immutable. 
On the contrary, different high-level languages could be mapped to the same, agreed, IR language. 
In QRmap, for instance, an easy way to produce the graph representation is by means of a graphical interface that automatically generates the IR.

Analyzing the code in Fig.~\ref{fig:intermediate} section by section, the first \textit{header section} reports the commands that define how to interpret the following code. 
In the example, the measurement unit \texttt{"m"} for distances (meter) and how to encode floating point numbers are specified. 
Command \texttt{ENCODE\_FLOAT 3 4} means that float numbers are encoded with $3$ bits for the exponent and $4$ bits for the mantissa. Allowing user customization of code reduces its size and increases the dimension of the program that can be stored in the sQRy.

The following \textit{label section} defines the three labels used in the example. Each label has a name (i.e., \texttt{common} for label 1, \texttt{arc} for label 2, and \texttt{ass\_line} for label 3)
and the description of all the attributes. 
For instance, for label 1 the first attribute \texttt{name} must be present, while attribute \texttt{desc} is optional. 
Instead, regarding label 3, the last attribute is associated with the measurement unit number 0, which corresponds to \texttt{"m"} as defined in the header session.

The \textit{graph section} defines the structure of the graph (both vertices and arcs) and the valorization of the attributes of associated labels. The graph is defined vertex by vertex, and each vertex is identified by the word \texttt{LABEL}. 
To save space, there is no explicit association of a numerical identifier to vertices, but they are numbered according to the order of their definitions in the graph section. 
Analyzing the first vertex that coincides with the emergency exit, it is associated with label 1. 
The keyword \texttt{ATTRIBUTES\_FLAGS} is followed by a Boolean value for each optional attribute, which identifies its presence. 
Then, the valorization of the two attributes (\texttt{"Emergency Exit"} and \texttt{"Use in ..."}) is reported, and finally, one or more commands of type \texttt{DESTINATION\_NODE} define the arcs exiting for the node. 
For instance, \texttt{DESTINATION\_NODE 2} defines the arc $e_{v_1 \leftrightarrow v_2}^{l_2}$, which is not oriented, and the label associated with the arc (i.e., label 2), followed by the valorization of the attributes.

In the \textit{quick choices section}, a quick choice is selected, which is aimed at guiding the user to destination node 1 using the path with minimal length, i.e., the attribute \texttt{"length"} of arcs is exploited to find the minimum path. 
Quick choices are typically rendered as separate buttons in the mobile applications executing the sQRy.

The final \textit{code section} makes use of a slightly modified version of the QRtree dialect \cite{10492995, QRtree-spec}, which is aimed at defining a decision tree, to request the user, with a set of questions, requirements about the destination and the path. 
Depending on the user feedback, the best search criteria are selected to guide the user to the intended destination. 
In this example, the program first asks the user to select the desired destination. 
If the answer is ``Assembly line'', the program asks the required production speed expressed in products per hour: if the inserted value is smaller or equal to $10$ the user is directed to assembly line 1 (vertex 5), otherwise she/he is directed to assembly line 2 (vertex 6).

For all these IR-level commands, conversion rules to the binary code and vice versa were defined. 
They are not listed here due to space limitations. 
As usual for sQRy, the primary goal in doing this conversion is to generate a binary code as compact as possible.
After applying these rules, the binary code obtained for the proposed example has an occupation of \SI{3241}{bits}, which corresponds to just $13.7$\% of the overall space available in a QR code. Since the main limitation of the technology is the storage capacity of QR codes, it can be partially addressed by adding in the path sQRy, containing partial information, to guide the user step by step to the destination.

\section{Conclusions}
\label{sec:conclusions}
A new QRmap dialect for embedding geographic maps into sQRy is first proposed and preliminarily defined in this work.
It allows embedding paths of a map as graphs in sQRy and, with extensible labels, it efficiently associates attributes with vertices and arcs. Additionally, it enables interaction with the user through a specific programming language based on decision trees, aimed at identifying the optimal path to the intended destination according to the specific user requirements.

The presented example shows a preliminary definition of the QRmap dialect to guide users in large industrial environments.
The implementation of the complete generation and execution chains, the exhaustive definition of the QRmap specification, and the experimentation with real human operators is left as future work.

\bibliographystyle{IEEEtran}
\bibliography{bibliography}

\cleardoublepage

\end{document}